\documentclass[groupedaddress,twocolumn,amsmath,amssymb,floatfix,pra]{revtex4}
\usepackage{graphicx}
\usepackage{dcolumn}
\usepackage{bm}

\begin{document}

\title{Adiabatic Quantum Computing in
systems with constant inter-qubit couplings}

\author{S. Knysh$^1$, V.N. Smelyanskiy$^2$}
\affiliation{$^1$Mission Critical Technologies, Inc., 2041
Rosecrans Avenue, Suite 225, El Segundo, CA  90245}
\affiliation{$^2$NASA Ames Research Center, Mail Stop 269-2,
Moffett Field, CA 94035, USA}

\begin{abstract}
We propose an approach suitable for solving NP-complete problems
via  adiabatic quantum computation with an architecture based on a
lattice of interacting spins (qubits) driven by locally adjustable
effective magnetic fields. Interactions between qubits are assumed constant and
instance-independent, programming is done only by changing local
magnetic fields. Implementations using qubits coupled by
magnetic-, electric-dipole and exchange interactions are
discussed.
\end{abstract}


\maketitle

\section{Introduction}

It is one of the central open questions in the field of quantum
computing if quantum algorithms can efficiently solve
computationally hard problems in combinatorial optimization.   The
most common
 problems of this kind encountered in practice
belong to a so-called NP-complete class \cite{Garey} and are  in
almost one-to-one correspondence with the spin glass models in
physics \cite{Parizi}. A basic property of these models is an
onset of an exponentially large number of deep local minima of the
energy landscape formed by spin configurations separated by a
large number of spin flips.

A general framework for solving optimization problems on a quantum
computer is provided by an Adiabatic Quantum Computation (AQC)
\cite{Farhi:00,FarhiSc}. In its simplest form  AQC corresponds to
a "quantum annealing" (QA) of a spin system in the uniform
transverse magnetic field described by a smoothly-varying in time
 Hamiltonian $\hat H$
\begin{equation}
\hat H(\Gamma)=-\Gamma\sum_{i=1}^{N}\hat\sigma_{i}^{x}+\hat
H_0,\quad \Gamma\equiv\Gamma(t).\label{H}
\end{equation}
\noindent  $\Gamma$ is proportional to the transverse field
strength; $\hat\sigma_{i}^{x}$ is a Pauli matrix for the $i^{\rm
th}$ spin. The first term in (\ref{H}) is called  a \lq\lq
driver", it  causes transitions between the eigenstates of the
problem Hamiltonian $\hat H_0$ whose ground state  encodes the
solution of the classical optimization problem in question. For
example, to find the ground state of the classical Ising model one
uses \begin{equation} \hat H_0=- \sum_{i=1}^{N} h_i \hat
\sigma_{z}^{i} - \sum_{\langle i, k \rangle} J_{ik} \,\hat
\sigma_{i}^{z} \hat \sigma_{k}^{z},\label{HP}
\end{equation}
\noindent where $h_i$ are longitudinal  fields acting on
 spins and the summation is over  the pairs $\langle i, k \rangle$ of
coupled lattice sites. In QA the value of $\Gamma=\Gamma(t)>0$ is
slowly decreasing in time from a  large value $\Gamma_0\gg
\max(|J_{ij}|)$ at the start, to $\Gamma=0$ at the end of the
algorithm. The initial state
 is prepared to be a ground state of
 $H(\Gamma_0)\approx -\Gamma_0\sum_{j=1}^{N}\hat\sigma_{j}^{x}$
 with each spin pointing in the positive x-direction.
For adiabatically slow variation of $\Gamma(t)$ the system state
will be closely tracking  the instantaneous ground state $
|\psi_0(\Gamma)\rangle$ of  $\hat H(\Gamma)$. It will approach the
ground state of $H_0$ at the end of the algorithm and  the
solution of the optimization problem can be recovered by
measurements.

A metric for the performance of the AQC can be given in terms of
the minimum value $g_{\rm min}=\min_{\Gamma}g(\Gamma)$ of the
excitation gap $g(\Gamma)$
 between the ground  and first excited
energy levels of $H(\Gamma)$. In particular, if $1/g_{\rm min}$
grows no faster then polynomially in the problem size $N$ then so
does the runtime of AQC.

It was demonstrated recently for AQC applied to  computationally
hard random instances of NP-complete Satisfiability problem that
the algorithm performance is substantially affected by the
existence of the first-order quantum phase transition for some
value of $\Gamma=\Gamma^*$ in the limit $N\rightarrow\infty$
\cite{skm:04}. In this limit the instantaneous gap  averaged over
the ensemble of random problem instances vanishes, $\langle
g(\Gamma^*)\rangle\rightarrow 0$. This produces a significant
difficulty in the analysis of the asymptotic complexity of AQC.
Indeed, assume  that for a given problem instance ${\cal I}_N$ of
a finite size $N$ the minimum gap is achieved at some value of
$\Gamma=\Gamma_{{\cal I}_N}$. For large $N$ the distribution of
the values of $\Gamma_{{\cal I}_N}$
 over the ensemble of  random problem instances ${\cal I}_N$ is peaked
around  $\Gamma_*$ and has a width $\sim N^{-1/2}$. Therefore the
minimum  of the ensemble-averaged excitation  gap,
$\min_{\Gamma}\langle g\rangle$, will scale down polynomially with
$N$. At the same time the ensemble average of the minimum gap $
\langle \min_{\Gamma} g(\Gamma)\rangle$ can be exponentially small
in $N$.

The above discussion implies that, in general, the order of
time-minimization (over $\Gamma$) and statistical averaging of the
excitation gap over the ensemble of problem instances
\emph{cannot} be inverted. Therefore static properties of the
quantum phase transition (e.g., a phase diagram) are not
sufficient to obtain the true  asymptotic complexity of AQC and a
daunting theoretical task of studying  the \emph{dynamics} of the
phase transition is required. An crucial insight here can be
gained from the experiments implementing AQC for problem instances
of a large size.

An early  work on this subject is   a   quantum annealing
experiment \cite{Brooke}. It uses the macroscopic samples of a
disordered magnet LiHo$_{\rm x}$Y$_{\rm 1-x}$F$_4$ with the
dipolar coupling between the spins that are formed by the doublet
states of Ho$^{+3}$ ions randomly substituted for nonmagnetic
Y$^{3+}$.  When the magnetic field  is applied perpendicular to
the Ising  axes the  system can be approximated by the Hamiltonian
$H$ (\ref{H}),(\ref{HP})  corresponding to a quantum 3D Ising
model with the random antiferromagnetic interaction between
neighboring spins  in the transverse magnetic field
$\Gamma=\Gamma(t)$.

For $\Gamma=0$ finding a ground state of $H$ can be mapped onto an
NP-complete problem \cite{Barahona:82}.
 However this mapping  is only approximate for the
random magnet \cite{Brooke} because it is not exactly of the Ising
type due to the very small components of the g-factor transverse
to the Ising axis. Also the  analog computation via QA
\cite{Brooke} if of a low-fidelity type because it does  not have
any explicit control nor correction of the
 decoherence-induced errors, unlike the  conventional quantum computation
 schemes based on quantum circuits.
 Finally,  the
adiabatic evolution in QA \cite{Brooke}
 collapses beyond the point $\Gamma_*>0$ of a quantum phase transition
from a quantum paramagnet to a quantum spin glass where the system
escapes  from the adiabatic ground state to states with low-lying
energy levels forming a quasi-continuous spectrum
\cite{Santore:02}.

Despite all these shortcomings  the magnetic susceptibility study
in \cite{Brooke} and the subsequent numerical simulations of this
system \cite{Santore:02} shows that QA provides a dramatic speedup
of convergence to low-energy states of LiHo$_{\rm x}$Y$_{\rm
1-x}$F$_4$ over  a purely thermal annealing procedure at zero
transverse magnetic field. This was attributed to the fact that
the tunnelling processes between local minima of the energy
landscape open up new transition pathways as compared to thermally
activated spin flips.

The decoherence rate for the lowest doublet (spin) states in Ho
ions comes from the coupling to Ho nuclei and is small
\cite{Giraud:01},\cite{Ghost}(a), possibly much smaller then the
level splitting and magnetic dipolar coupling between the
neighboring Ho spins \cite{Stamp:04}. One can expect that the
success of QA is based substantially
 on a strong  build-in quantum coherence with extremely entangled many-spin states
 and quantum correlations extending out over large
group of spins (cf. \cite{Ghost}(b)).

All NP-complete problems can be mapped  onto one another by a
classical  algorithms  that scale polynomially in the problem size
$N$ \cite{Garey}. However different problems correspond to spin
Hamiltonians with quite different physical properties, such as
long-range \emph{vs} short-range interactions, fully-connected
\emph{vs} sparsely-connected graphs,  etc. Therefore  it would be
of a fundamental importance to experimentally study the properties
of QA
 in mesoscopic systems modelling different
NP-complete problems at $\Gamma=0$. This can shed a light on the
power  of QA  even if the experiments are of a low-fidelity type
 \cite{Brooke}. However one has to overcome several obstacles to implement this
approach:
\begin{enumerate}
    \item[(i)]   many
NP-complete problems correspond to a long-range interaction
between  classical spins while the underlying physical systems
have short-range interactions;
    \item[(ii)] in most of the perspective material systems the
coupling between qubits is either  difficult or impossible to
control. The former includes systems with superconducting qubits
\cite{Makhlin:01,Mooij:99,Pashkin:03,Wellstood:03,Devoret_review:04}
and the latter  includes systems with magneto-dipole
\cite{Chudnovsky:01,Sousa:04,Loss_review:05}, electro-dipole
\cite{Platzman:99,Fedichkin:00,Wiel:03,Hollenberg:04}, and
 and
elastic-dipole \cite{Golding:03,spo:04} interactions;
    \item[(iii)] irregularities of the crystal lattice and random spatial positions of
     qubits give rise to the fluctuations
    in  magnitude and sign of the qubit coupling coefficients across the
    sample.
\end{enumerate}

 Kaminsky, Lloyd and Orlando
(KLO) \cite{klo:04} proposed a novel AQC architecture  for solving
NP-complete Maximum Independent Set problem using superconducting
qubits. In their method an instance of the problem is converted
into an equivalent circuit with ferromagnetic and
antiferromagnetic couplings in a uniform magnetic field. In
\cite{klo:04} switchable interqubit couplings using one of the
mechanisms presented in \cite{Mooij:99,jofet} are necessary for
encoding an instance of MIS, but the architecture does not require
a change in couplings during the algorithm execution for a given
problem instance.

At the same time the couplings in \cite{klo:04} have to be
readjusted for solving a new instance of MIS problem. Also the
scalability of this  is limited because it only allows the
switching between the regimes of small and large couplings. I.e.,
it cannot guarantee that two sites are completely decoupled, nor
the sign of the coupling can be changed.

In what following we propose the architecture for AQC that assumes
that qubit couplings are \emph{not} adjustable  and addresses all
the problems (i)-(iii) mentioned above. Moreover the architecture
that we propose for solving various NP-complete problems via AQC
is universal, in a sense that it is independent on any particular
qubit implementation.

\section{Maximum independent set problem}

For practical implementation of AQC for solving an optimization
problem it is necessary to translate this problem into a model
Hamiltonian with single-spin and pairwise interactions only. It is
especially straightforward to do for the NP-complete graph theory
problem  Maximum Independent Set (MIS) \cite{klo:04}. This problem
is defined on a graph $G=(V,E)$ that is composed of the set of
vertices $V$ connected by edges from the set $E$. MIS is the
problem of  finding the largest subset ${\cal S}$ of vertices $V$
such that no two vertices on the subset share an edge from $E$.
The reason why this problem is well suited form AQC is that its
solution is isomorphic to finding a ground state of an
antiferromagnetrically coupled Ising spin model in a uniform
magnetic field \cite{Barahona:82}.

\begin{figure}[h!]
\includegraphics[width=3in]{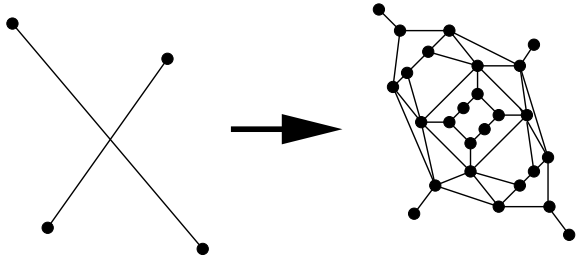}
\caption{Every intersection of two edges is replaced by a planar graph
(right). \label{fig:gadget}}
\end{figure}

The next obstacle has to do with the fact that we want to use
spins on a 2D lattice and use only nearest-neighbor interactions.
If the graph is non-planar it cannot be embedded into 2D lattice
with nearest-neighbor interactions only. And there is no guarantee
that a graph just constructed has a drawing on a plane without
intersecting edges. Fortunately using a construction invented by
Garey, Johnson and Stockmeyer \cite{garey:74} we can replace two
intersecting edges by a gadget of Fig.~\ref{fig:gadget}.
Maximizing the independent set in the gadget reveals opposite
vertices cannot both be occupied without incurring energy penalty.
Solving MIS problem for the transformed graph gives the solution
for the original one. Replacing all instances of intersecting
edges by such gadget renders the graph planar. For this reason,
without losing generality, we can restrict our discussion to
planar graphs.

Before we describe the embedding of planar graph into a regular 2D
lattice let us demonstrate how the MIS problem maps onto Ising
model. We model vertices of the graph with classical Ising spins
that can take values $s_i = \pm 1$. The edges correspond to
antiferromagnetic interactions between spins. Consider the
following classical Hamiltonian:
\begin{equation}
  H_0=-\sum_{\langle i,k \rangle} J_{i,k} s_i s_k + \sum_i s_i \sum_k J_{i,k}
  \label{eq:H0}
\end{equation}
where $J_{i,k}<0$ corresponding to antiferromagnetic interaction
and local magnetic fields are $h_i=\sum_k J_{i,k}$ (the sum is
over the neighbors of a vertex $i$). This expression can be (up to
a constant) rewritten as
\begin{equation}
  H_0=-\sum_{\langle i,k \rangle} J_{i,k} (1-s_i)(1-s_k).
\end{equation}
Since all interactions are antiferromagnetic ($J_{i,k}<0$) and
$(1-s_i)(1-s_k) \geq 0$, we must have $H_0 \geq 0$. The minimum
value $H_0=0$ is reached only if no edge $\langle i,k \rangle$ has
both $s_i=-1$ and $s_k=-1$. In other words the set of vertices
$\{i|s_i=-1\}$ is independent, no two vertices from this set share
a vertex. Hence, the minima of $H_0$ are all possible independent
sets. Next, we adjust individual magnetic fields $h_i'=h_i-J$,
which is  equivalent to adding a perturbation
\begin{equation}
  V=J \sum_i s_i.\label{perturbation}
\end{equation}
Its role is to maximize the number of vertices with $s_i=-1$, i.e.
maximize the cardinality of MIS. By choosing $J \leq \min_{\langle
i,k \rangle} |J_{i,k}|$ we guarantee that all minima of $H_0+V$
correspond to  independent sets with  $H_0=0$. Indeed, if for some
minimum of $H_0+V$ and for some $i,k$ both $s_i=-1$ and $s_k=-1$,
then setting $s_i=+1$ decreases $H_0$ by at least $4 |J_{i,k}|
\geq 4J$ and at the same time increases $V$ by $2J$. The overall
energy decreases by at least $2J$ contradicting our assumption
that we started from a minimum of $H_0+V$. Therefore, the ground
states of $H_0+V$ correspond to solutions of MIS problem.
Moreover, the gap between ground state(s) and first excited
state(s) is at least $2J$.

\begin{figure}[h!]
\includegraphics[width=3in]{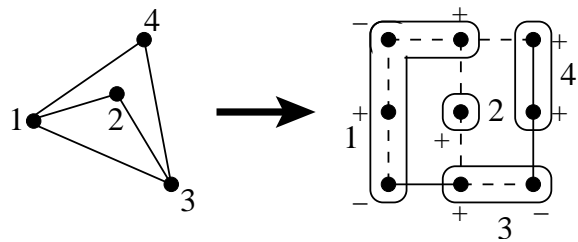}
\caption{An example graph with 4 vertices (left) is mapped onto another graph
with vertices lying on the grid (right). Each vertex is replaced by clusters
on the right. ``$+$'' and ``$-$'' signify values of
corresponding $\tau_{i\alpha}$;
solid and dashed lines represent ferromagnetic and antiferromagnetic
interactions respectively. \label{fig:clust}}
\end{figure}

Obviously, very few planar graphs can be drawn on a lattice, with
vertices corresponding to the nodes of the lattice and edges
corresponding to nearest-neighbor links. In general case vertices
have to be represented by whole clusters of spins. We generalize
Hamiltonian (\ref{eq:H0}) as follows. We label spins by two
indices; in addition to cluster index $i$, we have index within a
cluster $\alpha$:  $s_{i\alpha}$. Similarly, interactions are
denoted by $J_{i\alpha,k\beta}$ which can be either ferromagnetic
or antiferromagnetic. Restrictions on signs of interactions will
be explained later, while as before we require
$|J_{i\alpha,k\beta}| \geq J$. Within each cluster, the graph of
interactions is a connected tree. Between clusters there is at
most one link, and the link between clusters $i$ and $k$ is
present if and only if there is an edge joining vertices $i$ and
$k$. This construction is illustrated in Fig.~\ref{fig:clust} for
a simple graph.

The fact that the graph of interactions within a cluster is a tree means
that when considered separately, each cluster has a doubly degenerate
ground state. We identify them with values of fictitious coarse-grained
spin $S_i$. Value $S_i=-1$ shall indicate that vertex $i$ is included in
the independent set. The individual spins take values $s_i=\tau_i S_i$
for appropriate $\tau_i$.
Clearly
\begin{equation}
  J_{i\alpha,i\beta} \tau_{i\alpha} \tau_{i\beta} > 0. \label{ineq1}
\end{equation}

In principle, values of $\tau_{i\alpha}$ can be deduced from
$J_{i\alpha,i\beta}$ via the last equation, but in practice this
will not be necessary since their values will be evident from the
construction.

Between different clusters an interaction must be effectively
antiferromagnetic. If there is a link between $s_{i\alpha}$ and
$s_{k\beta}$, we require that
\begin{equation}
J_{i\alpha,k\beta} \tau_{i,\alpha} \tau_{k,\beta} < 0,
\label{ineq2}
\end{equation}
\noindent so that $S_i$ and $S_k$ are coupled
antiferromagnetically.

Consider the Hamiltonian
\begin{eqnarray}
   H_0 & = & -\sum_i \sum_{\langle i\alpha,i\beta \rangle}
                     J_{i\alpha,i\beta} s_{i\alpha} s_{i\beta}
             -\sum_{\langle i\alpha,k\beta \rangle}
                     J_{i\alpha,k\beta} s_{i\alpha} s_{k\beta} \nonumber \\
          && -\sum_{i\alpha}h_{i\alpha}s_{i\alpha}.\label{H0sum}
\end{eqnarray}
\noindent Here in the first term the double summation is over the
clusters $i$ and  over all edges $\langle i\alpha,i\beta\rangle$
 within each cluster.  In the second
term the summation is over all pairs of distinct clusters $i\neq
k$ connected by an edge $\langle i\alpha,k\beta\rangle$.
Individual fields $h_{i\alpha}$ in (\ref{H0sum}) can be adjusted
so that  $H_0$ is rewritten (up to a constant) as $H_0=H_1+H_2$
with
\begin{eqnarray}
H_1 & = & \sum_i \sum_{\langle i\alpha,i\beta \rangle}
|J_{i\alpha,i\beta}|
          (1-\tau_{i\alpha}\tau_{i\beta}s_{i\alpha}s_{i\beta}), \label{H1}\\
H_2 & = & \sum_{\langle i\alpha,k\beta \rangle} |J_{i\alpha,k\beta}|
          (1-\tau_{i\alpha}s_{i\alpha})(1-\tau_{k\beta}s_{k\beta}).\label{H2}
\end{eqnarray}\noindent
(it can be verified that all signs above are correct given the
inequalities (\ref{ineq1}) and (\ref{ineq2})). In the above
equations  both $H_1 \geq 0$ and $H_2 \geq 0$. The ground state of
$H_0$ corresponds to $H_1=0$ and $H_2=0$. The first equality
$H_1=0$ implies
$\tau_{i\alpha}s_{i\alpha}=\tau_{i\beta}s_{i\beta}$ for all
$\alpha$, $\beta$ or, equivalently,
$s_{i\alpha}=\tau_{i\alpha}S_i$ for all $\alpha$. With this
constraint in mind, $H_2$ can be rewritten into a familiar form
$\sum_{i\alpha,k\beta} |J_{i\alpha,k\beta}|(1-S_i)(1-S_k)$. Then
$H_2=0$ implies that no two vertices $i,k$ that are joined by an
edge belong to the independent set at the same time ($S_i=S_k=-1$
is forbidden).

We now estimate the gap between ground states and excited states.
If $H_1\neq 0$  it must be at least $2|J_{i\alpha,i\beta}|$; and
if $H_2\neq 0$ it is at least $4|J_{i\alpha,k\beta}|$. Therefore
 the gap is at least $2J$. The degenerate ground
states of $H_0$ correspond to all possible independent sets. This
degeneracy is (partially) lifted by adding a term to $H_0$
\begin{equation}
  H_0=H_1+H_2+V,\quad
  V=\frac{J}{2}\sum_i{s_{i0}\tau_{i0}},\label{H0V}
\end{equation}
where $0$ is some arbitrary index within a cluster. This
corresponds to adjusting $h_{i0}'=h_{i0}-\frac{J}{2}\tau_{i0}$.
Alternatively, rather than pick up a particular spin ${i\alpha}$
we can adjust
$h_{i\alpha}'=h'_{i\alpha}-\frac{J}{2n_i}\tau_{i\alpha}$, where
$n_i$ is the size of cluster $i$. Then
\begin{equation}
V=\frac{J}{2}\sum_{i\alpha}\frac{s_{i\alpha}\tau_{i\alpha}}{n_i}.\label{V1}
\end{equation}
\noindent In either case, the net effect is that the extra term
takes a form $\frac{J}{2} \sum_i S_i$, which favors independent
sets with the largest cardinality (number of vertices with
$S_i=-1$ is maximized). We need only to verify that the
perturbation $V$ is small enough to not mix the ground states and
excited states of $H_0$. If either $H_1$ or $H_2$ is positive,
then,  for all spins within the corresponding cluster $i$ we set
$s_{i\alpha}=\tau_{i\alpha}$ (this is equivalent to setting
$S_i=+1$). This decreases the energy by at least $2J$ as we have
already estimated. On the other hand, the energy increase due to
degeneracy-breaking term is at most $J$. We conclude that the
minima of $H_0$ involve only independent sets, and that the energy
gap is at least $J$.

Once we have constructed a classical Hamiltonian $H_0$
(\ref{H1})-(\ref{V1}) the ground state of which corresponds to the
solution of NP-complete MIS problem, we formulate a quantum
Hamiltonian $\hat H$ (\ref{H}) for use in the AQC by replacing in
$H_0$ all classical spins $s_{i\alpha}$ with operators
$\hat{\sigma}_{i\alpha}^z$ (cf. (\ref{HP})) and adding a driver
term $-\Gamma \sum_{i=1}^{N} \hat{\sigma}_{i}^x$.

\section{\label{sec:embedding} Embedding planar graph in a lattice}

As explained in the previous section, only planar graphs have to
be considered. Also note that only connected graphs can be
considered; for disconnected graphs we may solve the problem for
each connected component. Every vertex of the planar graph is
mapped onto a cluster of vertices and edges forming a tree. This
allows vertices far from each other to be joined. Also we must
always preserve an antiferromagnetic character of interactions
between the vertices to satisfy the inequality (\ref{ineq2}).

An arbitrary planar graph can be mapped onto an appropriately
sized regular lattice of spins with nearest neighbor interactions,
as long as such system has a frustration (loops with odd number of
antiferromagnetic couplings must exist). This is a necessary
condition since the ground state of a non-frustrated system is
trivially obtained, whereas a ground state of NP-complete problems
typically has an exponentially large degeneracy, with degenerate
states unrelated to each other by symmetry transformations.

For illustrative purposes we now embed a graph in a triangular
lattice with antiferromagnetic interactions (so that each
plaquette is frustrated), but such that every antiferromagnetic
coupling can be switched off by an external control. A
straightforward construction of a graph embedding is via iterated
procedure. Let's label the vertices with numbers $1 \ldots N$
subject to following constraints: a) vertex $i$ is necessarily
adjacent to at least one vertex $j$ such that $j<i$, and b) if a
set of vertices $j_1, j_2, \ldots j_n, i$ forms a face of the
planar graph and all $j_k < i$, all vertices that appear
\emph{inside} that face must have numbers smaller than $i$. Once
we constructed an embedding of subgraph $H$ of graph $G$, we
augment $H$ with a new vertex $i$ connected to it. We extend the
clusters corresponding to the vertices adjacent to $i$ and link
them to $i$. A completely blind procedure may not always work. Let
us call the  vertices from $H$ \lq\lq eligible'' if they are
adjacent to the vertices from $G$ that do not belong to $H$
($G\backslash H$). Links from eligible vertices to some new vertex
will have to be drawn at later stages.

\begin{figure}[h!]
\includegraphics[width=3in]{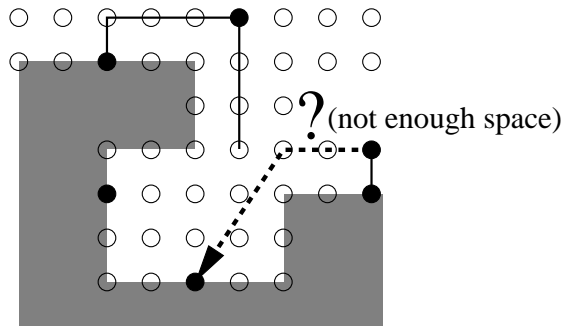}
\caption{An example of the case where  a graph's embedding has a
non-convex shape and some eligible vertices (black circles) lie
inside of the cavities. In this case there may not be enough space
to connect all eligible vertices to outlying vertices. A gray area
represents a part of the graph that is already embedded.
\label{fig:bottleneck}}
\end{figure}

\begin{figure}[h!]
\includegraphics[width=3in]{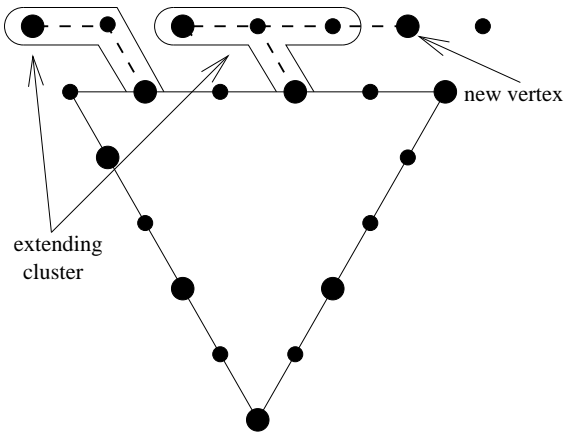}
\caption{All vertices on the top side are coupled
ferromagnetically to the new side. A new vertex is introduced that
is coupled antiferromagnetically \cite{explanation}.
\label{fig:em1}}
\end{figure}

\begin{figure}[h!]
\includegraphics[width=3in]{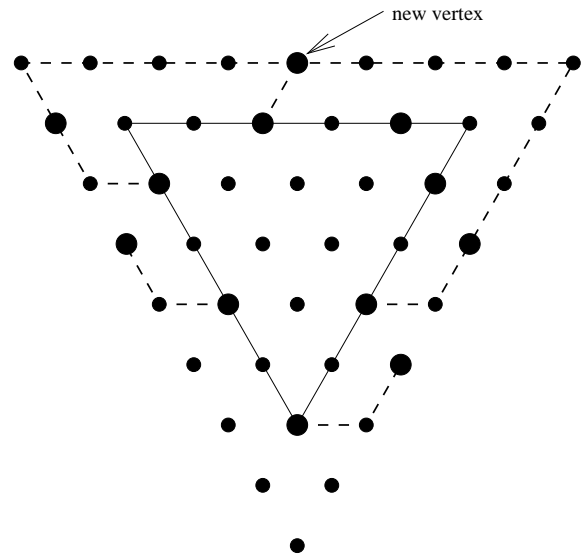}
\caption{A new vertex is connected via antiferromagnetic links.
All other vertices coupled ferromagnetically to new sides
\cite{explanation}. \label{fig:em2}}
\end{figure}

In the case of a non-convex  embedding of a subgraph $H$  we may
encounter a situation where there are not enough qubits in 2D
lattice to draw all links from the new vertex to eligible vertices
(see Fig.\ref{fig:bottleneck}). We can guarantee that this never
happens if we use the following guidelines: we always expand
clusters corresponding to eligible vertices so that the embedding
remains convex, eligible vertices lie on the perimeter of this
convex region, and a minimum distance between eligible vertices
along the perimeter is maintained.
\noindent
For the triangular lattice we make sure that the convex object
always has the form of a triangle. The distance between the
eligible vertices along the perimeter must be at least two. Adding
a vertex that is joined to subgraph $H$ by only one edge increases
the side of the triangle by 1 (see Fig.~\ref{fig:em1}), and adding
a vertex joined by two or more edges to subgraph $H$ can increase
the side of the triangle by 2 for the worst-case scenario depicted
in Fig.~\ref{fig:em2}.

We now explain Figs.~\ref{fig:em1} and \ref{fig:em2} in more
details. Large black circles lying on a boundary of the triangle
describe the eligible vertices; connections to them may have to be
made at a later stage. In Fig.~\ref{fig:em1} one vertex is added
joined to the current subgraphs by one edge. It must be coupled
antiferromagnetically to a cluster corresponding to neighboring
vertex \cite{explanation}.

 A \lq\lq copy'' of the neighboring vertex is made since
more vertices may connect to it later on. The cluster associated
with that vertex is expanded to include that copy. Similarly,
copies of all eligible vertices on the same side of the triangles
are added to corresponding clusters. Note that copies of vertices
are coupled ferromagnetically, whereas new vertex is coupled
antiferromagnetically. This guarantees that for all large black
circles the corresponding values of $\tau_{i\alpha}=1$ and the
inequalities (\ref{ineq1}),(\ref{ineq2}) are satisfied. In
Fig.~\ref{fig:em2} a new vertex is connected to 3 vertices of the
subgraph $H$. Copies of left and right vertices are made by
ferromagnetic couplings. No copy of the middle vertex is made --
it cannot be connected to new vertices as long as we add vertices
in correct order. Fortunately, under the iteration in both cases a
minimum distance of 2 between eligible vertices is maintained.

Let us now make some estimates. For a connected planar graph with
$N$ vertices and $M$ edges the following inequality holds:
\begin{equation}
   N-1 \leq M \leq 3(N-2).
\end{equation}
Due to this tight bound, $N$ alone is a good measure of the size of a graph.
Let us now estimate the number of qubits needed in our construction. Adding
each consecutive vertex may increase the side of the triangle by $2$.
A single vertex is represented by a single qubit. A triangle of size $2(N-1)$
has $N(2N-1) < 2N^2$ qubits. This presents an upper bound on the number of
qubits needed to model a planar graph with $N$ vertices. Note that we have
used qubits quite liberally. We hope that far smaller number of qubits may be
required for practical implementation, possibly a number scaling linearly
with $N$. Minimization of this number is an interesting topic for further
study.

\section{Adjustable couplings vs. fixed couplings}

The construction that we provided requires the programmability of
 nearest-neighbor interactions (i.e., the ability to switch them off).
 Embedding of NP-complete problems can only be done into the lattices that possess frustration.
An antiferromagnet on a triangular lattice  is always frustrated
and therefore only the ability to switch off antiferromagnetic
interactions is necessary in this case. However for a square
lattice, switching between ferromagnetic, antiferromagnetic and
zero couplings is necessary. This is because for a square lattice
neither ferromagnet nor antiferromagnet has a frustration that
requires the presence of interactions of both signs.

Antiferromagnetic triangular lattice with adjustable interactions
can be implemented in superconducting QC by one of devices
described in \cite{Mooij:99} and \cite{jofet}. The idea to use
them for implementing AQC was first put forward by Kaminsky, Lloyd
and Orlando \cite{klo:04}. However the approach described in
\cite{klo:04} cannot be used when switching between 3 states
(ferromagnetic, antiferromagnetic and zero coupling) is needed.
Moreover, the switching is not precisely between antiferromagnetic
and zero couplings but rather between strong antiferromagnetic and
weak antiferromagnetic couplings. While this is adequate for small
graphs, for graphs that are sufficiently large
($O(J_\textrm{on}/J_\textrm{off})$), presence of weak
antiferromagnetic coupling can lead to appearance of spurious
minima unrelated to the correct solution of MIS problem. This
effect can be compensated by grouping vertices in larger clusters
and grouping links between qubits in these clusters in larger
bundles. This approach is  analogous to using thick \lq\lq
wires'', which increases $J_\textrm{on}/J_\textrm{off}$ ratio for
effective interactions. This approach, however, is ridden with
complexities. And most importantly, as was argued in Introduction,
in many physical systems  it is crucial  to have an AQC
architecture that does not rely \emph{at all} on tunable
interactions. This is precisely what we propose to accomplish in
this paper.

\subsection{\label{fixed} Fixed interactions}

The general idea of our approach is the following. Instead of
\lq\lq deleting'' couplings between nodes of the regular lattice
we propose to delete vertices. By applying a large (effective)
magnetic field to a particular qubit $i$ we can polarize it along
the $z$ direction (to value $+1$). Magnetic fields for its
neighbors are adjusted as follows
\begin{equation}
  h_k'=h_k-J_{i,k}.
\end{equation}\noindent
The net effect is the same as if qubit $i$ was completely absent
(not coupled to its neighbors). Since this operation can be
applied to any qubit, we can start from a regular lattice and by
selecting a set of qubits to be deleted appropriately, we \lq\lq
carve out" a desired circuit. Note that, although we so far
assumed that spins at deleted vertices are completely polarized,
corresponding to infinitely large magnetic fields, in practice we
only require that spins be completely polarized in the ground
state of the classical Hamiltonian. For that purpose it is
sufficient to use magnetic fields on the order of $J$, defined as
$J=\min_{\langle i,k\rangle} J_{i,k}$.

\begin{figure}[h!]
\includegraphics[width=\linewidth]{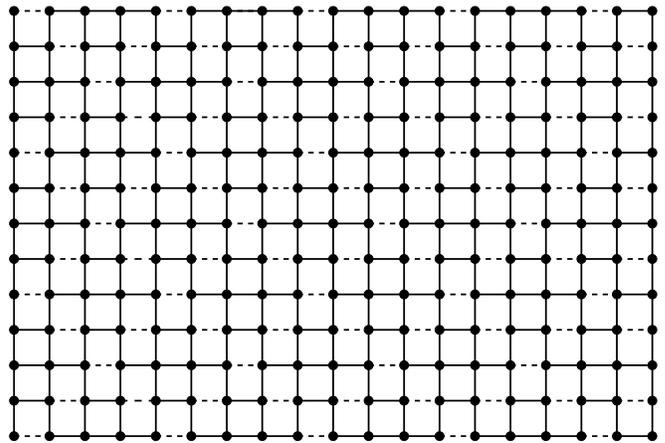}
\caption{A patterned lattice in which 50\% of elementary square loops are
frustrated \label{fig:old2}}
\end{figure}

We identify several solutions for implementing this. First general
approach is to use one lattice where individual nodes are \lq\lq
deleted" to simulate another lattice with switchable interactions.
The simulated lattice will have a larger lattice constant.
Therefore a larger number of qubits is required. As a first
example of this approach we take a square lattice with a
particular periodic structure as depicted in Fig.~\ref{fig:old2}.
In this figure solid lines correspond to ferromagnetic and dashed
line to antiferromagnetic couplings.
\noindent

\begin{figure}[h!]
\includegraphics[width=\linewidth]{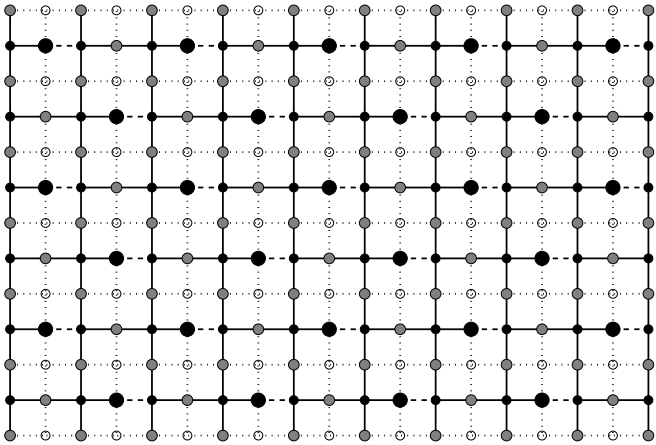}
\caption{Mapping onto triangular lattice with antiferromagnetic interactions.
Interactions are switched off by polarizing gray qubits. \label{fig:tri}}
\end{figure}

 In Fig.~\ref{fig:tri} we divide all qubits into four
sets. ``Deleted'' qubits are represented by empty circles; the
severed couplings are represented by dotted lines. ``Working''
qubits, depicted by large black circles correspond to the nodes of
triangular sublattice. Unused ``auxiliary'' qubits represented by
small black circles effectively ``pass on'' interaction to their
neighbors. The ``control'' qubits represented by gray circles are
used to selectively enable or disable antiferromagnetic couplings
between working qubits. Deleting gray control qubit switches off
an interaction between two working qubits that otherwise are
effectively coupled antiferromagnetically. As a next step we apply
our construction for embedding a planar graph into a triangular
lattice. Note that since for every working qubit we use 2
auxiliary, 3 control qubits and 2 deleted qubits, our estimate for
the number of qubits is correspondingly increased by a factor of
8.

\begin{figure}[h!]
\includegraphics[width=\linewidth]{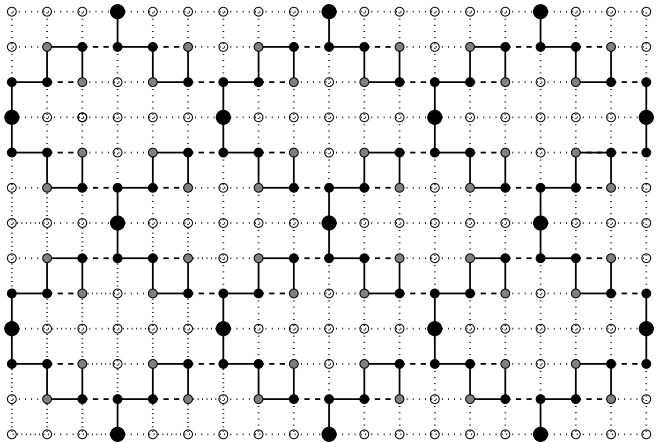}
\caption{Mapping onto a square lattice in which nearest neighbor
interactions can be made ferromagnetic, antiferromagnetic or can
be severed. \label{fig:sq}}
\end{figure}

For illustrative purposes we give another example shown in
Fig.~\ref{fig:sq}. This time a square lattice in which every
plaquette is frustrated is mapped onto another square lattice with
a lattice constant that is  $3\sqrt{2}$ times larger. Now the
control qubits can be used not only to switch off the interaction
but also to choose its sign. Couplings between a pair neighboring
working qubits is controlled by two control qubits. Deleting both
control vertices severs the interaction between the two working
qubits, while deleting only one makes an interaction effectively
ferromagnetic or antiferromagnetic, depending on which vertex was
deleted.

\begin{figure}[h!]
\includegraphics[width=3in]{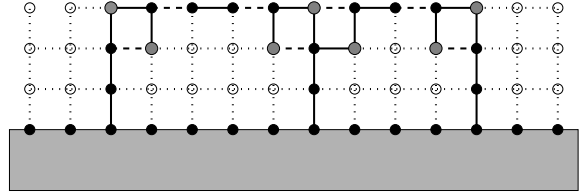}
\caption{In the center, either top gray qubit or two bottom gray
qubits are deleted. On the sides either of two gray qubits is
deleted. Since each elementary square is frustrated it is always
possible to make all links antiferromagnetic. A gray area
represents a part of the graph that is already embedded.
\label{fig:dir}}
\end{figure}

Yet another approach is to do graph embedding directly on a
lattice. We choose a particular square lattice with frustration
where each plaquette has 1 edge of one type (ferromagnetic or
antiferromagnetic) and 3 edges of another type. As before, we
construct the embedding interactively. Assuming that we
constructed an embedding for a subgraph of the original graph so
that all eligible vertices lie on the side of the square, we
couple the new vertex antiferromagnetically to some of these
eligible vertices. This can always be done since we are free to
reroute the connections at will. An example is shown in
Fig.~\ref{fig:dir}. Control qubits as before are represented by
gray circles. For left and right edges we delete either gray
qubit. For the center edge we either delete qubit 2 or both qubits
1 and 3. A simple counting argument shows that since there is a
total of 8 choices for control qubits, exactly one will correspond
to all couplings being effectively antiferromagnetic. This
procedure increases the side of the square by $6$ in the worst
case, taking the total number of qubits necessary to $36N^2$ in
the worst case. As before we stress that this is the worst-case
scenario and the estimate will be lowered if better layout
algorithms are used.

\section{Fault-tolerant computing}


For nanoscale computing architectures it is  important to take
into account irregularities of the crystal lattice.  For classical
processing units in molecular computing the following solution was
proposed \cite{terramac}. A massively parallel array of simple
computational devices is used, but in contrast to standard
architectures, interconnects between devices are redundant (note
that both devices and interconnects can be defective). The
presence of redundant connections allows to reroute the system
once defective elements are identified. Perfectly working system
was implemented despite the fact that 10\% of elements were
defective.

The present AQC architecture, in which individual qubits are
``deleted'' by polarizing  them is perfectly suited to address the
problem of fault-tolerant computation. In our model only a small
subset of nearest-neighbor couplings are utilized. From this
standpoint, the system has large redundancy of links. A computer
program, given database of defects, can achieve routings that
bypass defects altogether.

\begin{figure}[h!]
\includegraphics[width=3in]{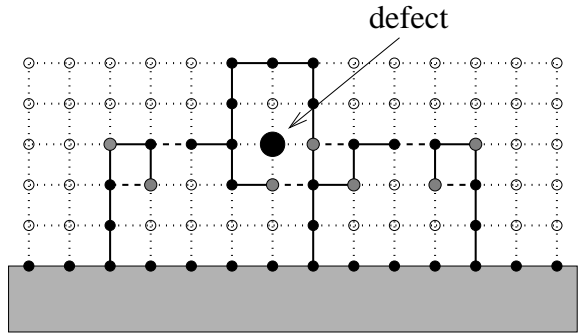}
\caption{If a it is always possible to route the links around
single defect as shown in this figure. A gray area represents a
part of the graph that is already embedded.\label{fig:def}}
\end{figure}

This can be easily illustrated as follows. Assume for simplicity
that only the following defects can occur: (i) a particular
nearest-neighbor coupling is too small in absolute value,
$|J_{i,k}| < J$ for some threshold $J$ (see
Eq.~(\ref{perturbation}) and discussion after it); (ii)  a
particular nearest-neighbor coupling has a wrong sign,
ferromagnetic when it should be antiferromagnetic or {\it vice
versa}. In either of those cases we mark one of the qubits as
``defective'', delete it and route all links around it as shown in
the Fig.~\ref{fig:def}. Similarly, we can route around qubits that
are themselves defective. This can be done as long as individual
defects are isolated and their concentration is not too big. In
general, it should be possible to reroute around the defects as
long as their concentration does not exceed the percolation
threshold.

Moreover, it is possible to do the routing algorithm even when dealing with
an entirely random pattern. The number of possible routings increases
exponentially with their length and with high probability we can always
find at least one routing that makes couplings effectively antiferromagnetic.
We expect that this can be done with the number of qubits that differs only
by a constant factor compared to the case of regular pattern.

\section{\label{sec:Implement} Implementation}

There exists a number of requirement for the implementations of
the above AQC architecture for solving NP-complete problems.
Firstly, the underlying  spin lattice must have frustration.
Secondly, an interaction between spins must be of the Ising type,
 $\propto \hat s_{i}^{z} \hat s_{j}^{z}$. Thirdly,  in our analysis above we assumed only
 a nearest-neighbor Ising coupling. The above requirements can be
 satisfied for qubits coupled via the  Heisenberg  exchange
 interaction,
 superconducting qubits, and, except for the last condition,
 in many systems with \lq\lq always on" magnetic-dipole and
electric-dipole interqubit coupling.

 To see this we consider the
Hamiltonian ${\cal H}_{12}$ for  magnetic coupling between the two
spins-1/2  located in the plane $xy$ and subject each to a strong
static  magnetic field $\hat z\, B_{\alpha}^{z}$ ($\alpha=1,2$).
Also each spin is resonantly driven by a weak ac magnetic field
$\vec B^{\rm ac}_\alpha(t)=\hat
x\,B_{\alpha}^{x}\cos(\Omega_\alpha t)$ applied in the orthogonal
direction. Assuming that the difference between the values of the
Zeeman splitting for the two spins is much greater then their
magnetic coupling we obtain in the rotating wave approximation
\cite{Slichter:96,Sousa:04}
\begin{equation}
{\cal H}_{12}=-\sum_{\alpha=1}^{2}\left(h_{\alpha} \hat
\sigma_{\alpha}^{z}+\Gamma \hat \sigma_{\alpha}^{x}\right)
+\left[D_{12}(d)+J(d)\right]\, \hat \sigma_{1}^{z}\, \hat
\sigma_{2}^{z},\label{H12}
\end{equation}
\noindent \vspace{-0.2in}
\begin{equation} h_{\alpha}=\frac{1}{2}\left(\hbar \Omega_\alpha-\gamma_\alpha\,
B_{\alpha}^{z}\right),\quad \Gamma=-\frac{1}{2}\gamma_\alpha
B_{\alpha}^{x},\nonumber
\end{equation}
\noindent \vspace{-0.2in}
\begin{equation}
D_{12}(d)=\frac{\hbar \gamma_1\gamma_2}{4d^3},\quad J(d)\propto
\exp(-d/a^*),\nonumber
\end{equation}
\noindent where
\begin{equation}
|h_{j}|,\,|\Gamma_j|\ll \Omega_j,\quad |D_{12}(d)|,\,|J(d)| \ll
\hbar|\Omega_{1}-\Omega_{2}|.\label{ineq}
\end{equation}
\noindent In (\ref{H12})  we  neglected small terms proportional
to $\hat s_{\pm}^{j}$ which amount to corrections quadratic in
$D_{12}/\hbar(\Omega_{1}-\Omega_{2})|$ and
$J/\hbar(\Omega_{1}-\Omega_{2})|$. The  coefficients $D_{12}(d)$
and $J(d)$  are magnetic-dipole and Heisenberg exchange constants,
respectively. Both of them are positive and correspond to
antiferromagnetic interaction between spins. Therefore if spins
are located at the vertices of a triangular lattice the later will
be frustrated.

By adjusting the amplitudes of the resonant driving fields
$B_{j}^{x}$ one can set the effective transverse field
$\Gamma_j=\Gamma$ for all spins except for the \lq\lq deleted"
spins where there is no ac driving ($\Gamma_j=0$).  Also each ac
field $\vec B^{\rm ac}_j(t)$ is not a sequence of short pulses but
rather a continues wave. During the AQC the field detuning from
resonance $h_j$ is fixed while the field matrix element
$\Gamma=\Gamma(t)$  is decreased adiabatically slow in time.
 This can be achieved using various  methods for a single spin qubit control, such as
modification of the electron g-factor \cite{Salis:01}, and others
that are discussed in the context of the spin-based QC in quantum
dots \cite{LD:98,LBD:00,Loss_review:05} and shallow donors
\cite{Sousa:04}.

Consider now a triangular lattice of charge qubits where each
qubit  $\alpha$ is encoded by the two orbital states with the wave
functions $\Psi_{n}^{\alpha}(x,y,z)$ ($n=0,1$) and the difference
between the energy levels equals $\Delta E_{j}$. We assume that
 the only nonzero
dipole matrix elements for each qubit correspond to a z-component
of the electric dipole vector, $p_{nm}^{j}=-e\langle
\Psi_{n}^{j}|z|\Psi_{m}^{j}\rangle$. Consider now a pair of qubits
\emph{1} and \emph{2} located in $xy$ plane on a distance $d$ from
each other and assume that the strength of the electro-dipole
interaction between them $D_{12}(d)\ll \Delta E_{1}-\Delta E_2$.
 Then the truncated 2-qubit interaction
Hamiltonian has the form ${\cal H}_{12}=D_{12}(d) \hat s_{1}^{z}
\hat s_{2}^{z}$, where $\hat s_{j}^{z}$ are z-components of
effective spin-1/2 operators acting on qubit states and
$D_{12}(d)=(p_{00}-p_{11})^2/(2\,\hbar\,d^3)>0$ is a constant of
an electro-dipole interaction (cf. \cite{Platzman:99}). This
interaction is of an \lq\lq antiferromagnetic" type and will lead
to frustration on a triangular lattice of qubits located in $xy$
plane.

In a usual QC setting  we assume that a control electrode is
located near each qubit and apply a time-modulation of the
potential of the local electrodes that produces  weak microwave
fields with the frequencies $\Omega_j\gg |\Omega_j-\Delta
E_{j}|/\hbar$. The z-component of each field will drive resonantly
the transition between the corresponding qubit states  in a
near-field regime. Then in the rotating wave approximation ${\cal
H}_{12}$ will take a form (\ref{H12}) ($J=0$) with  $h_j$ and
$\Gamma$ being, respectively, a detuning from the resonance and a
non-diagonal matrix element of the field for a qubit $j$. By
fixing $\{h_j\}$ and slowly reducing the value of $\Gamma$ to zero
one can implement the AQC architecture similarly to the previous
case. The example given above is relevant  for the charge qubits
encoded by electron states lithographically confined on a surface
of liquid helium \cite{Platzman:99}, and also  by laterally
quantized electron states in single-electron quantum wells in GaAs
heterostructures \cite{Wiel:03,Golding:03}.

A main limitation of the proposed architecture for systems with
dipole-dipole interactions is that it takes into account only
nearest-neighbor interqubit couplings. We note however that in our
approach each vertex of the graph in MIS problem is embedded in a
cluster with a large number of qubits. This picture may
effectively support the nearest-neighbor coupling approximation.
The detailed analysis of this problem will be done elsewhere.

Another  challenge of  the implementation schemes considered above
is that frequencies of resonant driving $\Omega_j$  must be
different for different qubits and therefore each qubit has to be
driven with its own ac field (in near-field regime). For frequency
range $\lesssim $ 10$^{12}$ Hz this can be done by modulating the
gate bias near each qubit. This condition can be satisfied for
electrons on helium \cite{Platzman:99}, for electronic states in
broad quantum dots \cite{Golding:03} and for spin qubit control
via the g-factor modulation \cite{LBD:00,Salis:01}.

 However for qubits encoded by  orbital states
of a shallow donors in GaAs \cite{Allen:05} with the long decay
time ($\sim$ 350ns) the intra-qubit frequencies $\Delta
E_{j}/\hbar$ are in the THz range and local ac driving is highly
problematic. One could consider in this case a global driving with
the  THz field containing a range of frequencies within the
spectral window that overlaps with the intra-qubits frequencies
$\{\Delta E_{j}/\hbar\}$ so that each qubit will be driven
resonantly by its own frequency component of the field. Fixed
detuning $h_j$  can be produced locally for each qubit using a
quadratic Stark effect from the field of the control electrode.

\section{Conclusion and Outlook of Future Work}

In the proposed AQC architecture   the interactions between qubits
are never turned off and never re-adjusted for solving any given
instance of an NP-complete problem. Moreover, the problem
instances can be embedded into the underlying Ising lattice even
in the presence of site imperfections and  random distribution of
the magnitudes of the Ising coupling coefficients. Also  in the
proposed scheme  there is no need to perform NMR-type refocusing
sequences of fast qubit gates that are used in conventional
quantum computing (QC) architectures \cite{Sousa:04,Pryadko:05}
and must satisfy the stringent conditions reflecting the multiple
time scales in the material system. In contrary, in our approach
the computation process is guided with the slow continues-time
variation of the parameters of single-qubit Hamiltonians avoiding
unwanted resonances with bulk excitations. These considerations
make our analog scalable AQC architecture especially advantageous
for solid state QC implementations that use ``always on''
magnetic- \cite{Chudnovsky:01,Sousa:04,Loss_review:05} and
electric-dipole
\cite{Platzman:99,Fedichkin:00,Wiel:03,Hollenberg:04,Golding:03,spo:04}
 interactions for entanglement generation. This architecture is also of interest
for QC implementations that use the Heisenberg exchange
interaction, such as electron spin qubits in quantum dots
\cite{LD:98,Loss_review:05}. It will allow to avoid an inter-qubit
\lq\lq J" gate, or any electrical control over the wavefunction
overlap, hence making a gate lithography much simpler and reducing
the sensitivity to electrode noise.

 For future work we leave the mapping of an
instance of the Maximum Independent Set problem onto an arbitrary
planar graph, edges of which represent the couplings between
qubits. It would allow us to work with completely random systems
that do not have any periodic structure. Though the problem of
deciding whether a certain instance can be mapped onto an
arbitrary planar graph is likely NP-complete, we only need to find
a good approximation algorithm, which can be designed to take
polynomial time and guarantee to find such a mapping  for
sufficiently large random planar graph with high probability.

We chose to implement a Maximum Independent Set problem because
the problem uniquely permits fluctuations in the magnitude of
$J_{i,k}$, as long as the sign is unchanged. For practical
problems involving the solution of Constraint Satisfaction problem
the problem is often first transformed into an instance of a
Maximum Independent Set. Although a mapping from Constraint
Satisfaction to Maximum Independent Set  exists, it may be
beneficial to work with a Constraint Satisfaction problem from the
outset. We have described earlier that every clause that involves
three variables can be replaced by the
 gadget in the form of planar graph. Mapping this
directly onto Ising model will reduce the number of extra vertices
and edges necessary. Maximum Independent Set problem limits our
possibilities as it assigns equal weight to every vertex that
appears in the independent set. Relaxing this constraint leads for
more efficient architectures for simulating a clause or for a
gadget that replaces crossing edges.

Recently Oliviera and Terhal developed a  mapping of an
\emph{arbitrary} quantum circuit onto a square lattice with
nearest neighbor interactions of three possible types: $\hat s_1^x
\hat s_2^x$, $\hat s_1^y \hat s_2^y$, $\hat s_1^z \hat s_2^z$, or
no coupling at all \cite{terhal}. This opens door for emulating an
arbitrary quantum computing algorithm as AQC with all spins on a
2D square lattice.

The construction done in \cite{terhal} is rigid: every quantum
circuit corresponds to a particular lattice. It is interesting to
explore if our approach of polarizing qubits to \lq\lq carve out''
circuits out of standard pattern can be applied in the case where
couplings $\hat s_1^x \hat s_2^x$ and $\hat s_1^y \hat s_2^y$
appear alongside $\hat s_1^z \hat s_2^z$, and to find out what
might be the optimal standard pattern. In addition it is
interesting to explore if the same can be accomplished starting
off from a completely random pattern. If a standard pattern is
sufficient (provided we can adjust fields acting on individual
qubits), it may be a very practical approach for solid-state
implementations of a universal quantum computer and could be
programmed to implement Shor's factoring algorithm.

\section{Acknowledgments}
We gratefully acknowledge  M.I. Dykman (Michigan State University)
for stimulating discussions. This work was supported by the
National Security Agency (NSA) and Advanced Research and
Development Activity (ARDA) under Army Research Office (ARO)
contract number ARDA-QC-P004-J132-Y05/LPS-FY2005.


\begin{thebibliography}{99}

\bibitem{Garey}M.R. Garey and D.S. Johnson, {\em Computers and Intractability. A Guide to the Theory
of NP-Completeness} (W.H. Freeman, New York, 1997)

\bibitem{Parizi} M. Mezard, G. Parizi, and M. Virasoro, {\em Spin glass theory and beyond}
(World Scientific, Singapore, 1987).

\bibitem{Farhi:00}   E. Farhi, et al.,
arXiv:quant-ph/0001106.

\bibitem{FarhiSc}  E. Farhi, J. Goldstone, S. Gutmann, J. Lapan, A. Lundgren, and D. Preda,
{\it Science} {\bf 292}, 472 (2001).

\bibitem{skm:04} V. N. Smelyanskiy, S. Knysh, and R. D. Morris Phys. Rev. E 70,
036702 (2004)

\bibitem{Brooke} J. Brooke, et al,  {\it Science},
v. 284, p. 779 (1999).

\bibitem{Barahona:82} F. Barahona, J. Phys. A: Math. Gen. {\bf 15}
3241 (1982).

\bibitem{Santore:02} G. E. Santoro, et al, {\it Science}, v. 295, p. 2447 (2002).

\bibitem{Giraud:01} R. Giraud, et al., Phys. Rev. Lett {\bf 87},
057203 (2001).

\bibitem{Ghost} (a) S. Ghost, et al., Science {\bf 296}, 2195
(2002); (b) S. Ghost, et al., Nature {\bf 425}, 48 (2002).

\bibitem{Stamp:04} P.C.E. Stamp and I.S. Tupistyn, Phys. Rev. B
 {\bf 69}, 014401 (2004).

\bibitem{Makhlin:01} Y. Makhlin, G. Sch\"on, and A. Shnirman,
Rev. Mod. Phys. {\bf 73} 357 (2001).

\bibitem{Mooij:99} J.E. Mooij, et al.,  Science {\bf 285} 1036 (1999).

\bibitem{Pashkin:03} Yu. A. Pashkin, et al., Nature {\bf 421},
823 (2003).

\bibitem{Wellstood:03} A.J. Berkley, et al., Science {\bf 300}, 1548 (2003).

\bibitem{Devoret_review:04} M.H. Devoret, A. Wallraff and J.M. Martinis, cond-mat/0411174.

\bibitem{Platzman:99}  (a) P.M. Platzman and M.I. Dykman,  {\it Science} {\bf 284}
1967 (1999); (b) M.I. Dykman M I and P.M. Platzman, Fortschr. {\bf
48}, 1095 (2000).

\bibitem{Fedichkin:00} L. Fedichkin, M. Yanchenko and K.A. Valiev,
Nanotechnology {\bf 11}, 387 (2000); L. Fedichkin and A. Fedorov
Phys. Rev. A 69, 032311 (2004).

\bibitem{Wiel:03} W.G. Van der Wiel W G,et al.,  Rev. Mod. Phys. {\bf 75},
1 (2003).

\bibitem{Hollenberg:04} L. C. L. Hollenberg, et al.,
Phys. Rev. B 69, 113301 (2004).

\bibitem{Golding:03}  B. Golding, M.I. Dykman, cond-mat/0309147.

\bibitem{spo:04} V. N. Smelyanskiy, A. G. Petukhov, V. V. Osipov,
Phys. Rev. B RC 72, 081304 (2005); see also  quant-ph/0407220.


\bibitem{Chudnovsky:01} J. Tejada, et al, Nanotechnology {\bf 12},
181 (2001).

\bibitem{Sousa:04}R. de Sousa, J. D. Delgado, and S. Das Sarma
Phys. Rev. A 70, 052304 (2004).

\bibitem{Loss_review:05} V. Cerletti, et al, acrXiv:cond-mat/0412028.

\bibitem{klo:04} W. M. Kaminsky, S. Lloyd, T. P. Orlando, \lq\lq Scalable Architecture
for Adiabatic Quantum Computing of NP-Hard Problems", in  {\em
Quantum Computing $\&$ Quantum Bits in Mesoscopic Systems} (Kluwer
Academic 2003).


\bibitem{garey:74}
M. R. Garey, D. S. Johnson, L. Stockmeyer,   Proceedings of the
sixth annual ACM symposium on Theory of computing, p.47 (1974).

\bibitem{jofet}
M.J.~Storcz and F.K.~Wilhelm, Appl. Phys. Lett. 83, 2387 (2003).


\bibitem{explanation} Throughout the paper we refer to the
coupling between spins as being \lq\lq effectively" ferromagnetic
or antiferromagnetic if the chain of edges of the graph
connecting the spins contains even (odd) number of
antiferromagentic links.



\bibitem{terramac}
J.R.~Heath, P.J.~Kuekes, G.S.~Snider, R.S.~Williams, Science {\bf
280}, 1717 (1998).


\bibitem{LD:98} D. Loss and D.P. DiVincenzo, Phys. Rev. A {\bf
57}, 120 (1998).

\bibitem{Slichter:96} C.P. Slichter, {\it Principles of Magnetic
Resonance}, 3rd Ed. (Springer-Vergal, Berlin, 1996).

\bibitem{Salis:01} G. Salis, et al, Nature {\bf 414}, 619 (2001).

\bibitem{LBD:00} D. Loss, G. Burkard and D. P. DiVincenzo, Journal  of
Nanoparticle Research {\bf 2}, 401 (2000).

\bibitem{Allen:05} D.G. Allen, C.R. Stanley and M.S. Sherwin, arXiv:quant-ph/0503056.

\bibitem{Pryadko:05}P. Sengupta and L. P. Pryadko
Phys. Rev. Lett. 95, 037202 (2005).


\bibitem{terhal} R.~Oliveira, B.~Terhal,
arXiv:quant-ph/0504050.


\end{thebibliography}
\end{document}